\documentclass[aps,twocolumn]{revtex4-1}
\usepackage[colorlinks, linkcolor=red,anchorcolor=green,citecolor=blue]{hyperref}
\usepackage{graphicx}
\usepackage{bm}
\usepackage{amsmath}
\usepackage{color}
\usepackage{multirow}

\begin{document}
\title{Incomplete electromagnetic response of hot QCD matter}
\author{Zeyan Wang$^a$}
\author{Jiaxing Zhao$^a$}
\author{Carsten Greiner$^b$}
\author{Zhe Xu$^a$}
\author{Pengfei Zhuang$^a$}
\affiliation{$^a$Department of Physics, Tsinghua University, Beijing 100084, China}
\affiliation{$^b$Institut f\"ur Theoretische Physik, Johann Wolfgang Goethe-Universit\"at Frankfurt, Max-von-Laue-Strasse 1, 60438 Frankfurt am Main, Germany}
\date{\today}
\begin{abstract}
The electromagnetic response of hot QCD matter to decaying external magnetic
fields is investigated. We examine the validity of Ohm's law and find that the induced electric
current increases from zero and relaxes towards the value from Ohm's law. The relaxation time
is larger than the lifetime of the external magnetic field for the QCD matter in relativistic
heavy-ion collisions. The lower than expected electric current significantly suppresses
the induced magnetic field and makes the electromagnetic response incomplete.
We demonstrate the incomplete electromagnetic response of hot QCD matter by calculations
employing the parton transport model combined with the solution of Maxwell's equations. 
Our results show a strong suppression by two orders of magnitude in the magnetic field,
relatively to calculations assuming the validity of Ohm's law. This may undermine experimental
efforts to measure magnetic-field-related effects in heavy-ion collisions.
\end{abstract}

\maketitle

Noncentral relativistic heavy-ion collisions can generate a very strong magnetic field. 
Theoretical studies showed that its maximum value can reach 
$5m_\pi^2\approx 10^{18}\text{G}$ in Au+Au collisions at the top energy of RHIC and
almost $70m_\pi^2\approx 10^{19}\text{G}$ in Pb+Pb collisions at LHC 
energies~\cite{Skokov:2009qp,Voronyuk:2011jd,Deng:2012pc,Tuchin:2013ie},
where $m_\pi$ is the pion mass. The symmetry and phase structure of quantum
chromodynamics (QCD) are dramatically affected by strong magnetic fields~\cite{Andersen:2014xxa,Kharzeev:2012ph,Bali:2011qj,Bali:2012zg,Bruckmann:2013oba,DElia:2021yvk}. Experimentally, the strong electromagnetic fields give rise to searching for fantastic phenomena 
such as the chiral magnetic effect~\cite{Fukushima:2008xe,Kharzeev:2007jp,Kharzeev:2015znc,Zhao:2019hta}.
Furthermore, electromagnetic field-related phenomena such as the splitting of $D$ and
$\bar D$ directed flow~\cite{Das:2016cwd,Chatterjee:2018lsx,Zhao:2020jqu,STAR:2019clv,Acharya:2019ijj}, 
spin polarized difference between $\Lambda$ and $\bar \Lambda$~\cite{STAR:2017ckg,Muller:2018ibh,Guo:2019mgh,Guo:2019joy}, 
and photon involved QED (quantum electrodynamics) and QCD processes~\cite{ALICE:2015mzu,Zha:2018ytv,Klein:2018fmp,STAR:2019wlg,Brandenburg:2021lnj} 
have been intensively investigated by experimental and theoretical scientists.
 
However, it is hard to find the fingerprint of the electromagnetic fields from the measured
observables. This may lie on the weak signal of spin-related quantum fluctuations or
the short lifetime of the electromagnetic fields. How the electromagnetic fields evolve in 
relativistic heavy-ion collisions is a basic question for studying electromagnetic effects
on the QCD matter and thus, attracts great attention and broad interests~\cite{Gursoy:2014aka,Tuchin:2015oka,Zakharov:2014dia,Pu:2016ayh,Inghirami:2016iru,Yan:2021zjc,Stewart:2017zsu,Stewart:2021mjz,Chen:2021nxs}.

The essential issue is the electromagnetic response of the QCD matter to the fast decay of
the external electromagnetic field caused by the spectators.  A Faraday current ${\bf j}$ will
induce a magnetic field in the same direction as the external magnetic field, which prolongs
the lifetime of the total magnetic field, known as Lenz's law. Usually, people used Ohm's law
${\bf j}^{Ohm}=\sigma_{el} ( {\textbf{\emph{E}}}+{\textbf{\emph{V}}}\times {\textbf{\emph{B}}})$
to calculate the induced magnetic field. $\sigma_{el}$ is the electrical conductivity and
${\textbf{\emph{V}}}$ is the hydrodynamic velocity of the QGP \cite{Deng:2012pc}. 
However, the buildup of ${\bf j}$ will need some time to relax to ${\bf j}^{Ohm}$.
During this period, the induced magnetic field cannot be so large as expected so far
\cite{Gursoy:2014aka,Tuchin:2013ie}. We call this as incomplete electromagnetic response,
which has been overlooked in earlier studies. 

The incomplete electromagnetic response might be unremarkable for matter with large spatial
and timescales. The QCD matter produced in heavy-ion collisions is very special because
of its small spatial and time scale. The lifetime of the external electromagnetic field is
$R_A/\gamma \lesssim 0.06 \mbox{ fm/c}$ at top RHIC and LHC energies~\cite{Deng:2012pc,Tuchin:2013ie,Hattori:2016emy}.
The average creation time of QGP is about $0.6 \mbox{ fm/c}$ as utilized in numerous studies~\cite{Schenke:2010nt,Kestin:2008bh}. If the relaxation time of ${\bf j}$ for reaching
Ohm's law, denoted as $\tau_j$, is comparable with or even larger than these timescales,
the incomplete electromagnetic response will become significant.

Before we present numerical calculations to demonstrate the incomplete electromagnetic 
response of QCD matter, we estimate the relaxation time $\tau_j$ of the electric current
in $\phi$ direction on the $x-z$ plane with the help of the Drude model. This electric current
induces a magnetic field in the $y$ direction. According to the Drude model,
the change of particle momentum is caused by the Lorentz force and collisions with other
particles. The Lorentz force accelerates the collective motion, while collisions resist
the acceleration. The counterbalance will lead to the electric current obeying Ohm's law. 
The generation of the electric current is a non-Markovian process. During $dt$, the probability
of a collision is $dt/\tau_m$, where $\tau_m$ is the mean time between two subsequent
collisions. For a particle with momentum $p$, the change after a collision is a fraction of
$p$, $\Delta p \approx -p/\alpha$ on average. Then,  the mean change of momentum of
a quark in $\phi$ direction on the $x-z$ plane is 
\begin{eqnarray}
d\langle p_{\phi} \rangle(t)=-\frac{dt}{\tau_m \alpha} \langle p_{\phi}\rangle(t)+q_iE_{\phi}dt \,,
\label{eq.momentum1}
\end{eqnarray}
where $q_i$ is the electric charge of the quark species. We neglect the contribution of  
${\textbf{\emph{V}}}\times {\textbf{\emph{B}}}$, since it is much smaller than 
${\textbf{\emph{E}}}$ near midrapidity. We have also 
$j_\phi^{Ohm} \approx \sigma_{el} E_\phi$. The electric current is approximately
$j_\phi\approx\sum_i q_i n_i \langle p_\phi\rangle/\langle E \rangle$, where $n_i$ is
the particle number density and  the averaged energy is $\langle E \rangle=3T$.
Equation (\ref{eq.momentum1}) is rewritten in the form
\begin{eqnarray}
\frac{d}{dt} \frac{j_\phi(t)}{j_\phi^{Ohm}}=-\frac{1}{\tau_m \alpha} 
\frac{j_\phi(t)}{j_\phi^{Ohm}}+\frac{\sum_i q_i^2 n_i}{3T \sigma_{el}} \,.
\label{eq.current}
\end{eqnarray}
Suppose the solution has a relaxation form
\begin{eqnarray}
\frac{j_\phi(t)}{j_\phi^{Ohm}}=1- e^{-t/\tau_j} \,.
\label{eq.solution}
\end{eqnarray}
We have then
\begin{equation}
\label{relaxtime}
\tau_j=\tau_m \alpha =\frac{3T \sigma_{el}}{\sum_i q_i^2 n_i} \sim \sigma_{el}/T^2\,.
\end{equation}

The electrical conductivity of QCD matter has been studied by perturbative QCD~\cite{Arnold:2003zc},
lattice QCD~\cite{Gupta:2003zh,Aarts:2007wj,Buividovich:2010tn,Ding:2010ga,Burnier:2012ts,Brandt:2012jc,Amato:2013naa},
effective models~\cite{Finazzo:2013efa,Sahoo:2018dxn}, as well as transport
approaches~\cite{Cassing:2013iz,Greif:2014oia,Puglisi:2014sha,Steinert:2013fza}.
Their results of $\sigma_{el}/T$ is between $0.001$ and $0.4$ and thus differ greatly 
by two orders of magnitude, see also Ref. \cite{Zhao:2020jqu}.
For a typical choice of $\sigma_{el}/T= 0.03$ and temperature $T=255 \mbox{ MeV }$
of the QGP~\cite{Gursoy:2014aka}, Eq. (\ref{relaxtime}) gives $\tau_j = 1.12 \mbox{ fm/c}$
for two-quark flavors. This time scale is larger than the lifetime of the external 
electromagnetic field as well as the formation time of the QGP. We expect that the incomplete
electromagnetic response of the QGP will be significant and cannot be ignored.
In the following we demonstrate this incomplete electromagnetic response of QCD matter
within kinetic transport calculations.

The space-time evolution of quarks and gluons in the presence of electromagnetic fields
can be expressed by the relativistic Boltzmann transport equation,
\begin{eqnarray}
p^\mu\partial_\mu f_i + K^{\mu}{\partial \over \partial p^\mu}f_i =  {\mathcal{C}}[f_i],
\label{eq.Boltzmann}
\end{eqnarray}
where $K^\mu\equiv q_iF^{\mu \nu}p_\nu =(p_0{\boldsymbol{v}}\cdot{\bf F}_{\text{Loz.}} , p_0{\bf F}_{\text{Loz.}})$
is four-vector Mikowski force. $F^{\mu \nu}$ is the electromagnetic field strength tensor
and ${\bf F}_{\text{Loz.}}=q_i({\boldsymbol{v}}\times {\textbf{\emph{B}}}+{\textbf{\emph{E}}})$
is the Lorentz force. ${\boldsymbol{v}}={\boldsymbol{p}}/p_0$ is the particle velocity.
${\mathcal{C}}[f_i]$ stands for the collision term.

The total electromagnetic fields are the sum of the external fields and the one from the
responding medium. The latter is also the sum over all the electromagnetic fields from
moving quarks. For a particular quark as a source, the electromagnetic fields from it
can be obtained from the Li{\'e}nard-Wiechert potential
\begin{eqnarray}
\begin{split}
e{\textbf{\emph{E}}}_i({\textbf{\emph{r}}},t)=\frac{e^2}{4\pi}q_i
\Bigg(
\frac{{\textbf{\emph{n}}_s}-{\boldsymbol{\beta}}_s}
{{\gamma}^2(1-{\boldsymbol{\beta}}_s\cdot\textbf{\emph{n}}_s)^3
{\left|{\textbf{\emph{r}}}-{\textbf{\emph{r}}}_s\right|}^2}\\
+\frac
{\textbf{\emph{n}}_s \times \big( (\textbf{\emph{n}}_s-{\boldsymbol{\beta}}_s) \times \dot{{\boldsymbol{\beta}}_s}\big)}
{(1-{\boldsymbol{\beta}}_s\cdot\textbf{\emph{n}}_s)^3
{\left|{\textbf{\emph{r}}}-{\textbf{\emph{r}}}_s\right|}}
\Bigg)_{tr},
\label{eq.potentialE}
\end{split}
\end{eqnarray}
\begin{eqnarray}
\begin{split}
e{\textbf{\emph{B}}}_i({\textbf{\emph{r}}},t)=\frac{e^2}{4\pi}q_i
\Bigg(
\frac{{\boldsymbol{\beta}}_s\times{\textbf{\emph{n}}_s}}
{{\gamma}^2(1-{\boldsymbol{\beta}}_s\cdot\textbf{\emph{n}}_s)^3
{\left|{\textbf{\emph{r}}}-{\textbf{\emph{r}}}_s\right|}^2}\\
+\frac
{\textbf{\emph{n}}_s \times \Big( \textbf{\emph{n}}_s \times \big( (\textbf{\emph{n}}_s-{\boldsymbol{\beta}}_s) \times \dot{{\boldsymbol{\beta}}_s}\big)\Big)}
{(1-{\boldsymbol{\beta}}_s\cdot\textbf{\emph{n}}_s)^3
{\left|{\textbf{\emph{r}}}-{\textbf{\emph{r}}}_s\right|}}
\Bigg)_{tr},
\label{eq.potentialB}
\end{split}
\end{eqnarray}
where ``tr'' means that except for ${\bf r}$ and $t$, the other quantities are
evaluated at the retarded time $t_s=t-{\left|{\textbf{\emph{r}}}-{\textbf{\emph{r}}}_s\right|}$.
$\textbf{\emph{r}}_s$ is the particle position at $t_s$.
${\textbf{\emph{n}}}_s=({\textbf{\emph{r}}}-{\textbf{\emph{r}}}_s)/|{\textbf{\emph{r}}}-{\textbf{\emph{r}}}_s|$ denotes the unit vector pointing in the direction from the source.
${\boldsymbol{\beta}_s}={\boldsymbol{v}}(t_s)/c$, $\gamma$ is the Lorentz factor and $\dot{{\boldsymbol{\beta}}_s}$ is the derivative with respect to $t$.

Since the mass of $u$ and $d$ quarks is $m_q=3 \mbox{ MeV}$, $v$ is nearly equal to $c$.
At certain space-time points, the electromagnetic fields are huge for tiny
$(1-{\boldsymbol{\beta}}_s\cdot\textbf{\emph{n}}_s)^3$ in the denominators
in Eqs. (\ref{eq.potentialE}) and (\ref{eq.potentialB}). Therefore, large fluctuations of
electromagnetic fields will appear in one single heavy-ion collision. By averaging
a huge number of collision events one could obtain a visible response of QGP to
the decaying external fields. For numerical calculations in this work, we make
an approximation to take out fluctuations of the electromagnetic fields, as we replace
the velocity of the moving quark ${\boldsymbol{v}}$ in Eqs. (\ref{eq.potentialE}) and
(\ref{eq.potentialB}) by ${\bar {\boldsymbol{v}}}$,  the local average velocity of the same
species of quark as the source quark. By comparing the approximate and the exact
results that are the convolution of Eqs. (\ref{eq.potentialE}) and (\ref{eq.potentialB})
with a thermal distribution function, we find that the approximation will lower the
electromagnetic fields by a factor of $1- 1.5$ at $T=0.255 \text{ GeV}$ depending
on values of ${\bar {\boldsymbol{v}}}$ and ${\textbf{\emph{r}}}$.

Collisions of quarks and gluons are calculated by employing the parton cascade
model Boltzmann Approach of MultiParton Scatterings (BAMPS) \cite{Xu:2004mz},
in which the scattering probabilities in each spatial cell and at each time step, derived from 
the collision term,  are treated in a stochastic way and the test particle method is used to
enhance the statistics. In addition, the change of momentum of quarks due to 
the Lorentz force is done within each time step \cite{Greif:2017irh}.
 
Although the electrical conductivity was calculated within BAMPS including all binary
and radiation processes of partons \cite{Greif:2014oia}, in this work we assume binary
collisions with constant isotropic cross section. In this case, there is a simple relationship 
between the electrical conductivity and the total cross section according to the relaxation-time
approximation~\cite{Greif:2014oia,Puglisi:2014sha,Cassing:2013iz,Steinert:2013fza},
namely, 
\begin{eqnarray}
\sigma_{el}= {1\over 3T} \tau \sum_i q_i^2n_i \,,
\label{eq.electric.conductivity}
\end{eqnarray}
where $\tau$ is the relaxation time for a slight deviation from equilibrium. Comparing
Eq. (\ref{eq.electric.conductivity}) with Eq. (\ref{relaxtime}) we obtain $\tau_j=\tau$.
For binary collisions with constant isotropic cross sections we have 
$\tau=1/\sum_i n_i\sigma_{t}$ with $\sigma_{t}=2\sigma_{22}/3$ \cite{Xu:2007aa}.
$\sigma_{22}$ is the total cross section.
By varying the total cross section we can see different electromagnetic response to
the external fields. Qualitatively, we see from Eq. (\ref{eq.electric.conductivity})
that the electrical conductivity is inversely proportional to the total cross section. 
The smaller the cross section, the larger is the electrical conductivity and thus, the 
more significant would be the electromagnetic response. On the other hand, the smaller
the cross section, the larger is the relaxation time for the induced electric
current to reach Ohm's law. The electromagnetic response will be suppressed.
In the following we present quantitative results from numerical calculations. 

For our study we assume equilibrium states of gluons and quarks with $u$, $d$ two flavors.
Quarks and gluons are distributed uniformly in coordinate space and by the Boltzmann
distribution function in momentum space. 
The temperature of the system is set to be $T=255 \mbox{ MeV}$.  These setups lead to
$\sigma_{el}=11.6/(\sigma_{22}/\mbox{mb}) \mbox{ MeV}$. 
We use $\sigma_{22}=1$ and $2 \mbox{ mb}$,
which correspond to $\sigma_{el}=11.6$ and $5.8 \mbox{ MeV}$. 
In these cases, the shear viscosity to entropy density ratio
\cite{Hirano:2005wx,El:2012cr} is $\eta/s=0.44$ and $0.22$, respectively.

At first, we consider a linearly attenuated magnetic field as the external field,
$e\textbf{\emph{B}}_{ex}(t)=(0.1-0.01t/{\text{fm}}) \, {\bf e}_z \, \text{GeV}^2$,
which induces an electric field  
$e\textbf{\emph{E}}(r)=(0.005r/{\text{fm}})\, {\bf e}_{\phi} \, \text{GeV}^2$.
Such fields can be generated from an infinitely long solenoid by 
linearly decreasing the electric current twined around it. In our simulation, we 
consider a cylinder with a radius of $3 \mbox{ fm}$ and a length of $6  \mbox{ fm}$.
The thermal quark-gluon system is embedded in the cylinder and initialized at 
time $0 \mbox{ fm/c}$. A periodic boundary condition is taken in the $z$ direction,
while particles are reflected at the wall in the $r$ direction. For this condition,
the collective motion can only be in the $\phi$ direction. The radial motion due to
the Lorentz force and the reflection from the wall cancel each other out.

\begin{figure}[!tb]
\includegraphics[width=0.48\textwidth]{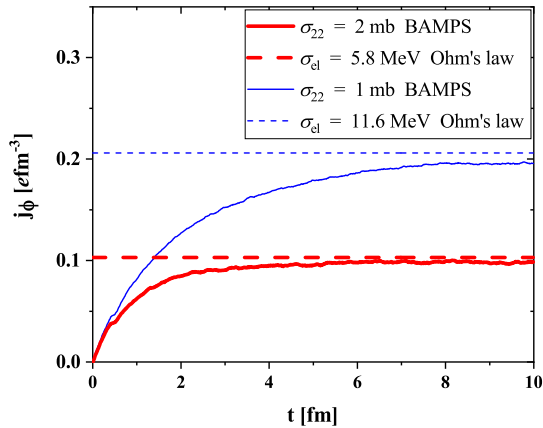}
\caption{The electric current density along the $\phi$ direction at $r=2.5 \, \text{fm}$.
Solid curves are the results from BAMPS calculations, while the dashed lines are
the results from Ohm's law $j_{\phi}^{Ohm}=\sigma_{el} E_{\phi}$. The total
cross section is set to be $1$ or $2 \mbox{ mb}$.}
\label{fig.constantE}
\end{figure}

The induced electric field generates an electric current. The electric current density at
a radius $r$ is calculated by
\begin{eqnarray}
j_\phi(t)={1\over V}\sum_i q_i{p_\phi^i(t)\over p_0^i(t)},
\label{eq.current}
\end{eqnarray}
where the summation is over all quarks in the cylindrical shell within radius 
$[r-\Delta r/2: r+\Delta r/2]$. $V$ is the volume.
The solid curves in Fig.~\ref{fig.constantE} show the buildup of the electric current density
taken at $r=2.5 \mbox{ fm}$ with $\Delta r = 0.1 \mbox{ fm}$. 
%---------------------------------------------------------------------
%---------------------------------------------------------------------
We see that the induced electric field  generates a circular electric current, which
is increasing from zero and relaxing towards the value from Ohm's law.
Also, the relaxation time with $\sigma_{22}=1 \mbox{ mb}$ is longer than that with 
$\sigma_{22}=2 \mbox{ mb}$, as expected. Even though, the generated electric
current with $\sigma_{22}=1 \mbox{ mb}$ is always larger than that with
$\sigma_{22}=2 \mbox{ mb}$. The results in Fig.~\ref{fig.constantE} can actually be
understood by the Drude model as discussed before. For $\sigma_{22}=1, 2 \mbox{ mb}$,
the relaxation time is $\tau_j=1.72,  0.86\mbox{ fm/c}$, respectively, according to
Eqs. (\ref{relaxtime}) and (\ref{eq.electric.conductivity}). These values are just $10\%$
smaller than the numerical results.

\begin{figure}[!tb]
\includegraphics[width=0.48\textwidth]{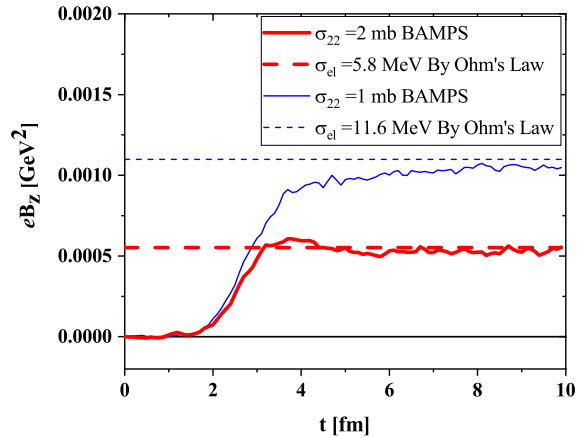}
\caption{The induced magnetic field at ${\bf r}=  0$.
The solid curves are the results from
BAMPS calculations, while the dashed lines are the results with Ohm's law.}
\label{fig.linearB}
\end{figure}

The electric current will generate a magnetic field denoted as the induced magnetic field, 
which should counteract the decay of the external magnetic field obeying Lenz's law.
In Fig.~\ref{fig.linearB} we show the generation of the induced magnetic field at ${\bf r}=0$
by the solid curves. The dashed lines represent the results calculated with Ohm's law.
Similar to the electric current, the induced magnetic field is increasing from zero and relaxing
towards the values calculated with Ohm's law. The induced magnetic field at early time
$t < 3 \mbox{ fm/c}$ seems independent of the total cross section or the electrical conductivity
of QGP. The reason may lie in the retardation of the electromagnetic fields.
%---------------------------------------------------------------------
%-------------------------------------------------------------------------

From Figs.~\ref{fig.constantE} and \ref{fig.linearB} we see the agreements of the 
BAMPS results at late times with those from or with Ohm's law, even a little lower 
because of the finite length of the cylinder. These indicate that the approximation we
made for the calculation of the electromagnetic fields of moving quarks is applicable.
At this point, the example with a linearly decreasing external
magnetic field serves as  a cross-check on the additional numerical implementations.

\begin{figure}[!t]
\includegraphics[width=0.48\textwidth]{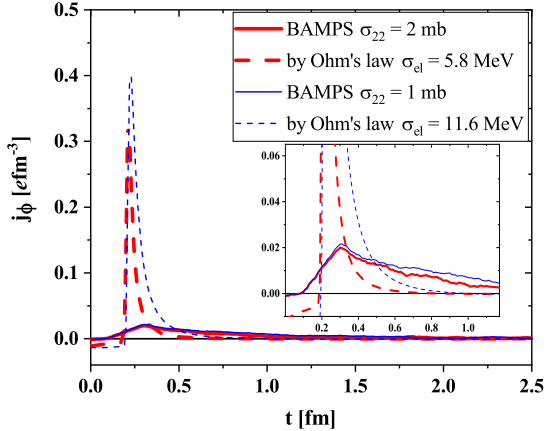}
\caption{The electric current density at $(x=2\,\text{fm}, y=0, z=0.2 \, \text{fm})$.
The solid curves are the results from BAMPS, while the dashed curves are the results by 
solving Eqs.~(\ref{eq.max1}) and (\ref{eq.max2}) and using Ohm's law.}
\label{fig.dmj}
\end{figure}

\begin{figure}[!t]
\includegraphics[width=0.48\textwidth]{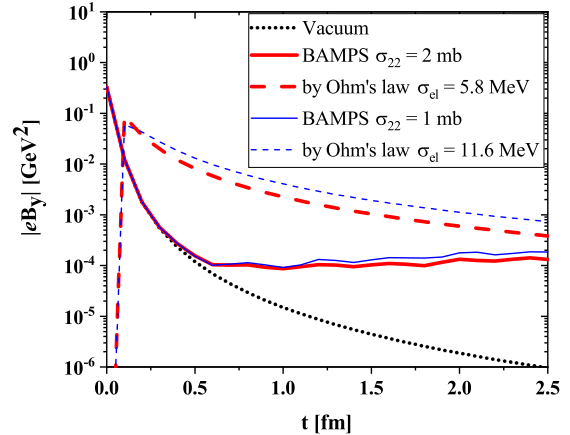}
\caption{The magnetic field at ${\bf r} = 0 \mbox{ fm}$.
The dotted curve depicts the external field,  the solid curves are the total
magnetic field (induced plus external field) from BAMPS calculation, and the dashed
curves are the total magnetic field by solving Eqs.~(\ref{eq.max1}) and (\ref{eq.max2}). }
\label{fig.dmB}
\end{figure}

Now we turn to another assumption of the external magnetic field, which mimics 
the situation in noncentral relativistic heavy-ion collisions and has been considered in
Ref. \cite{Tuchin:2013ie}. In this paper two nuclei are replaced by two
point particles with the same charge and mass (for a more realistic consideration see
Ref. \cite{Gursoy:2014aka}). They are moving in the $z$ direction at
impact parameter ${\bf b}=b \ {\bf e}_x$ and generate external electromagnetic fields
for a quark-gluon system, which is assumed to be static, thermal and
fill the whole coordinate space. The electromagnetic fields of the quark-gluon system are 
solved by Maxwell's equations
\begin{eqnarray}
\begin{split}
{\nabla}\cdot\textbf{\emph{B}}=0,
{\nabla}\times\textbf{\emph{E}}=-\frac{\partial\textbf{\emph{B}}}{\partial t},
\label{eq.max1}
\end{split}
\end{eqnarray}
\begin{eqnarray}
\begin{split}
{\nabla}\cdot\textbf{\emph{E}}&=q\delta(x-b/2)\delta(y)\delta(z-vt) \\
&+q\delta(x+b/2)\delta(y)\delta(z+vt)\,, \\
{\nabla}\times\textbf{\emph{B}}&=\frac{\partial\textbf{\emph{E}}}{\partial t}
+\sigma_{el}\textbf{\emph{E}} \\
&+qv{\hat{\textbf{z}}}\delta(x-b/2)\delta(y)\delta(z-vt) \\
&-qv{\hat{\textbf{z}}}\delta(x+b/2)\delta(y)\delta(z+vt)\,.
\label{eq.max2}
\end{split}
\end{eqnarray}
$v$ denotes the velocity of the nucleus, which is $v=(1-4m_N^2/s)^{1/2}$.
$m_N$ is the mass of a nucleon and $\sqrt{s}$ is the colliding energy per nucleon pair.
In Refs. \cite{Tuchin:2013ie,Gursoy:2014aka}, 
the ${\textbf{\emph{V}}}\times {\textbf{\emph{B}}}$ term in Ohm's law was not taken into account.
Recall that  ${\textbf{\emph{V}}}$ is the hydrodynamic velocity of the QCD matter. We repeat
the calculations in Ref. \cite{Tuchin:2013ie} and show results for 
$q=79e$, $\sqrt{s}=200 \mbox{ GeV}$, $b=7.6 \mbox{ fm}$, and $\sigma_{el}=5.8$ and 
$11.6 \mbox{ MeV}$ in Figs.~\ref{fig.dmj} and \ref{fig.dmB} by dashed curves.

%---------------------------------------------------------------------
%-------------------------------------------------------------------------

In our calculation with BAMPS we assume that a static and thermal quark-gluon system
appears at $t=0 \mbox{ fm/c}$, when two point-like nuclei are at the closest distance.
Furthermore, the quark-gluon system is embedded in a cube with a length of 
$6~ \text{fm}$ between the two nuclei. Periodic boundary conditions are used.
The temperature of the quark-gluon system is set to be $T=255 \,\text{MeV}$ and
the total cross section is $\sigma_{22}=2$ or $1 \mbox{ mb}$,
which corresponds to $\sigma_{el}=5.8$ or $11.6\, \text{MeV}$.
The external electromagnetic fields are calculated via Eqs. (\ref{eq.potentialE}) and
(\ref{eq.potentialB}) for moving point-like nuclei. 

When two point-like nuclei are approaching each other, the electromagnetic fields at
the central region around ${\bf r}=0 \mbox{ fm}$ are increasing. When the two point-like
nuclei are at the closest distance (collision at $t=0 \mbox{ fm/c}$), the magnetic field is 
at the maximum. Therefore, the induced electric field disappears at this moment and
is increasing again (but in opposite direction), when two point-like nuclei are departing
from each other.

Figure~\ref{fig.dmj} shows the electric current density at 
$(x=2 \mbox{ fm}, y=0, z=0.2 \mbox{ fm})$, generated by the total electric field.
We plot the $\phi$-component of the current in polar coordinate system 
on the $x-z$ plane, since it is responsible for the induced magnetic field in the $y$ direction. 
The electric field has a strong $x$ component. Thus, $E_\phi$ is almost $-E_x z/|x|$. 
The solid curves are the results from BAMPS calculation, while
the dashed ones are obtained by solving Eqs.~(\ref{eq.max1}) and (\ref{eq.max2}) 
and using Ohm's law.
We see promptly increasing electric currents when explicitly using Ohm's law.
The current with $\sigma_{el}=5.8 \mbox{ MeV}$ peaks within the time window 
$[0.2:0.4 \mbox{ fm/c}]$, while that with $\sigma_{el}=11.6 \mbox{ MeV}$ peaks within
$[0.2:0.6 \mbox{ fm/c}]$. Compared with these, the results from BAMPS show almost
twentyfold lower peaks with longer tails. The large difference is due to the relaxation of
the electric current towards Ohm's law.

%---------------------------------------------------------------------
%---------------------------------------------------------------------

The external magnetic field is increasing before $t=0\mbox{ fm/c}$. Thus, the induced
magnetic field is in the opposite direction as the external field.
By passing $t=0\mbox{ fm/c}$, the induced electric field reverses. If the quark-gluon
system exists before $0 \mbox{ fm/c}$, the induced magnetic field reverses too, but
with delay due to the retardation. Therefore, after $0\mbox{ fm/c}$ the total magnetic
field by solving Eqs.~(\ref{eq.max1}) and (\ref{eq.max2}) is smaller than the external
magnetic field for a while. This is shown in Fig.~\ref{fig.dmB} by the dashed curves for
the total magnetic field at ${\bf r}=0 \mbox{  fm}$.
The dotted curve depicts the external field.
By explicitly using Ohm's law, the electromagnetic response of the quark-gluon system
is much more significant. At $t=0.5 \mbox{ fm/c}$ for instance, the induced magnetic field
is larger than the external field by two orders of magnitude, although it is ten times smaller
than the maximal value of the external field at $t=0 \mbox{ fm/c}$.

In the calculation with BAMPS, the induced magnetic field is generated after
$t=0 \mbox{ fm/c}$. The result demonstrates the incomplete electromagnetic response.
The induced magnetic field is tiny at early times. Only after $t= 0.5 \mbox{ fm/c}$ does
it become dominant over the external field and stay almost constant. However,
the induced magnetic field after $0.5 \mbox{ fm/c}$ is smaller than the maximal field at 
$0\mbox{ fm/c}$ by three orders of magnitude. Such a small value may suppress any
magnetic effects. Besides the incompleteness, one has to take into account
the retardation of the generation of the magnetic field.
From Fig.~\ref{fig.dmB} we also see that the magnetic field is not sensitive to
the cross section, especially from $0$ to $0.5 \mbox{ fm/c}$.
Even though the application of kinetic transport models for QCD medium with
large coupling (large cross sections) may be in question, we do not expect a strong
increase of the magnetic field after $0.5 \mbox{ fm/c}$ in real heavy-ion collisions.

Some discussions are in order. The quark-gluon system we considered is an idealized 
one. In real heavy-ion collisions the produced quark-gluon system expands, cools, and
hadronizes. The temperature decreases and the electrical conductivity of the QGP changes
with time. Although all these will also influence the electromagnetic response of the QGP to 
the external fields, its incompleteness we proposed makes an important contribution and
should be taken seriously into account. Moreover, at the early stage of heavy-ion collisions,
the produced quarks and gluons are far from kinetic and chemical equilibrium~\cite{Blaizot:2011xf,Xu:2004mz,Romatschke:2017vte}. 
Quark number is smaller than that at thermal equilibrium, which will lead to a smaller induced 
electrical current. All of these considerations can be included in a more comprehensive study
by using BAMPS with more realistic initial condition and pQCD interactions. The results will
be presented in a future presentation.

In summary, in this Letter we have challenged the validity of Ohm's law applied for the hot
QCD matter produced at the early stage of noncentral relativistic heavy-ion collisions.
We have proposed the incomplete electromagnetic response due to the relaxation of
the induced electric current towards Ohm's law. The incomplete electromagnetic response
of hot QCD matter to external electromagnetic fields has been demonstrated by using
the parton cascade model BAMPS combined with the solution of Maxwell's equations.
The numerical results showed the significant suppression of the induced magnetic field.
This makes the search for any magnetic effects in heavy-ion experiments even more
challenging.

%{\bf Acknowledgement:} 
This work was financially supported by the National Natural Science Foundation of China
under Grants No. 11890710, No. 11890712, No. 12035006, and No. 12047535.
C.G. acknowledges support by the Deutsche Forschungsgemeinschaft (DFG) through
the grant CRC-TR 211 ``Strong-interaction matter under extreme conditions.''
The BAMPS simulations were performed at Tsinghua National Laboratory for Information
Science and Technology and on TianHe-1(A) at National Supercomputer Center in Tianjin.  
%--------------------------------------------------------------------------------------------------------------------------------


\begin{thebibliography}{20}

%\cite{Skokov:2009qp}
\bibitem{Skokov:2009qp}
V.~Skokov, A.~Y.~Illarionov and V.~Toneev,
%``Estimate of the magnetic field strength in heavy-ion collisions,''
Int. J. Mod. Phys. A \textbf{24}, 5925-5932 (2009).
%doi:10.1142/S0217751X09047570
%[arXiv:0907.1396 [nucl-th]].
%768 citations counted in INSPIRE as of 05 Dec 2020

%\cite{Voronyuk:2011jd}
\bibitem{Voronyuk:2011jd}
V.~Voronyuk, V.~D.~Toneev, W.~Cassing, E.~L.~Bratkovskaya, V.~P.~Konchakovski and S.~A.~Voloshin,
%``(Electro-)Magnetic field evolution in relativistic heavy-ion collisions,''
Phys. Rev. C \textbf{83}, 054911 (2011).
%doi:10.1103/PhysRevC.83.054911
%[arXiv:1103.4239 [nucl-th]].
%312 citations counted in INSPIRE as of 09 Dec 2020

%\cite{Deng:2012pc}
\bibitem{Deng:2012pc}
W.~T.~Deng and X.~G.~Huang,
%``Event-by-event generation of electromagnetic fields in heavy-ion collisions,''
Phys. Rev. C \textbf{85}, 044907 (2012).
%doi:10.1103/PhysRevC.85.044907
%[arXiv:1201.5108 [nucl-th]].
%425 citations counted in INSPIRE as of 05 Dec 2020

%\cite{Tuchin:2013ie}
\bibitem{Tuchin:2013ie}
K.~Tuchin,
%``Particle production in strong electromagnetic fields in relativistic heavy-ion collisions,''
Adv. High Energy Phys. \textbf{2013}, 490495 (2013).
%doi:10.1155/2013/490495
%[arXiv:1301.0099 [hep-ph]].
%263 citations counted in INSPIRE as of 07 Dec 2020

%\cite{Andersen:2014xxa}
\bibitem{Andersen:2014xxa}
J.~O.~Andersen, W.~R.~Naylor and A.~Tranberg,
%``Phase diagram of QCD in a magnetic field: A review,''
Rev. Mod. Phys. \textbf{88}, 025001 (2016).
%doi:10.1103/RevModPhys.88.025001
%[arXiv:1411.7176 [hep-ph]].
%228 citations counted in INSPIRE as of 05 Dec 2020

%\cite{Kharzeev:2012ph}
\bibitem{Kharzeev:2012ph}
D.~E.~Kharzeev, K.~Landsteiner, A.~Schmitt and H.~U.~Yee,
%``'Strongly interacting matter in magnetic fields': an overview,''
Lect. Notes Phys. \textbf{871}, 1-11 (2013).
%doi:10.1007/978-3-642-37305-3\_1
%[arXiv:1211.6245 [hep-ph]].
%192 citations counted in INSPIRE as of 05 Dec 2020

%\cite{Bali:2011qj}
\bibitem{Bali:2011qj}
G.~S.~Bali, F.~Bruckmann, G.~Endrodi, Z.~Fodor, S.~D.~Katz, S.~Krieg,
A.~Schafer and K.~K.~Szabo,
%``The QCD phase diagram for external magnetic fields,''
JHEP \textbf{02}, 044 (2012).
%doi:10.1007/JHEP02(2012)044
%[arXiv:1111.4956 [hep-lat]].
%582 citations counted in INSPIRE as of 16 Feb 2022

%\cite{Bali:2012zg}
\bibitem{Bali:2012zg}
G.~S.~Bali, F.~Bruckmann, G.~Endrodi, Z.~Fodor, S.~D.~Katz and A.~Schafer,
%``QCD quark condensate in external magnetic fields,''
Phys. Rev. D \textbf{86}, 071502 (2012).
%doi:10.1103/PhysRevD.86.071502
%[arXiv:1206.4205 [hep-lat]].
%432 citations counted in INSPIRE as of 16 Feb 2022

%\cite{Bruckmann:2013oba}
\bibitem{Bruckmann:2013oba}
F.~Bruckmann, G.~Endrodi and T.~G.~Kovacs,
%``Inverse magnetic catalysis and the Polyakov loop,''
JHEP \textbf{04}, 112 (2013).
%doi:10.1007/JHEP04(2013)112
%[arXiv:1303.3972 [hep-lat]].
%252 citations counted in INSPIRE as of 16 Feb 2022

%\cite{DElia:2021yvk}
\bibitem{DElia:2021yvk}
M.~D'Elia, L.~Maio, F.~Sanfilippo and A.~Stanzione,
%``Phase diagram of QCD in a magnetic background,''
Phys. Rev. D \textbf{105}, no.3, 034511 (2022).
%doi:10.1103/PhysRevD.105.034511
%[arXiv:2111.11237 [hep-lat]].
%4 citations counted in INSPIRE as of 27 Mar 2022

%\cite{Fukushima:2008xe}
\bibitem{Fukushima:2008xe}
K.~Fukushima, D.~E.~Kharzeev and H.~J.~Warringa,
%``The Chiral Magnetic Effect,''
Phys. Rev. D \textbf{78}, 074033 (2008).
%doi:10.1103/PhysRevD.78.074033
%[arXiv:0808.3382 [hep-ph]].
%1425 citations counted in INSPIRE as of 05 Dec 2020

%\cite{Kharzeev:2007jp}
\bibitem{Kharzeev:2007jp}
D.~E.~Kharzeev, L.~D.~McLerran and H.~J.~Warringa,
%``The Effects of topological charge change in heavy ion collisions: 'Event by event P and CP violation',''
Nucl. Phys. A \textbf{803}, 227-253 (2008).
%doi:10.1016/j.nuclphysa.2008.02.298
%[arXiv:0711.0950 [hep-ph]].
%1500 citations counted in INSPIRE as of 28 Sep 2021


%\cite{Kharzeev:2015znc}
\bibitem{Kharzeev:2015znc}
D.~E.~Kharzeev, J.~Liao, S.~A.~Voloshin and G.~Wang,
%``Chiral magnetic and vortical effects in high-energy nuclear collisions\textemdash{}A status report,''
Prog. Part. Nucl. Phys. \textbf{88}, 1-28 (2016).
%doi:10.1016/j.ppnp.2016.01.001
%[arXiv:1511.04050 [hep-ph]].
%485 citations counted in INSPIRE as of 28 Sep 2021

%\cite{Zhao:2019hta}
\bibitem{Zhao:2019hta}
J.~Zhao and F.~Wang,
%``Experimental searches for the chiral magnetic effect in heavy-ion collisions,''
Prog. Part. Nucl. Phys. \textbf{107}, 200-236 (2019).
%doi:10.1016/j.ppnp.2019.05.001
%[arXiv:1906.11413 [nucl-ex]].
%47 citations counted in INSPIRE as of 28 Sep 2021

%\cite{Das:2016cwd}
\bibitem{Das:2016cwd}
S.~K.~Das, S.~Plumari, S.~Chatterjee, J.~Alam, F.~Scardina and V.~Greco,
%``Directed Flow of Charm Quarks as a Witness of the Initial Strong Magnetic Field in Ultra-Relativistic Heavy Ion Collisions,''
Phys. Lett. B \textbf{768}, 260-264 (2017).
%doi:10.1016/j.physletb.2017.02.046
%[arXiv:1608.02231 [nucl-th]].
%97 citations counted in INSPIRE as of 05 Dec 2020

%\cite{Chatterjee:2018lsx}
\bibitem{Chatterjee:2018lsx}
S.~Chatterjee and P.~Bozek,
%``Interplay of drag by hot matter and electromagnetic force on the directed flow of heavy quarks,''
Phys. Lett. B \textbf{798}, 134955 (2019).
%doi:10.1016/j.physletb.2019.134955
%[arXiv:1804.04893 [nucl-th]].
%34 citations counted in INSPIRE as of 28 Nov 2021

%\cite{Zhao:2020jqu}
\bibitem{Zhao:2020jqu}
J.~Zhao, K.~Zhou, S.~Chen and P.~Zhuang,
%``Heavy flavors under extreme conditions in high energy nuclear collisions,''
Prog. Part. Nucl. Phys. \textbf{114}, 103801 (2020).
%doi:10.1016/j.ppnp.2020.103801
%[arXiv:2005.08277 [nucl-th]].
%3 citations counted in INSPIRE as of 12 Sep 2020

%\cite{STAR:2019clv}
\bibitem{STAR:2019clv}
J.~Adam \textit{et al.} [STAR],
%``First Observation of the Directed Flow of $D^{0}$ and $\overline{D^0}$ in Au+Au Collisions at $\sqrt{s_{\rm NN}}$ = 200 GeV,''
Phys. Rev. Lett. \textbf{123}, no.16, 162301 (2019).
%doi:10.1103/PhysRevLett.123.162301
%[arXiv:1905.02052 [nucl-ex]].
%51 citations counted in INSPIRE as of 28 Nov 2021

%\cite{Acharya:2019ijj}
\bibitem{Acharya:2019ijj}
S.~Acharya \textit{et al.} [ALICE],
%``Probing the effects of strong electromagnetic fields with charge-dependent directed flow in Pb-Pb collisions at the LHC,''
Phys. Rev. Lett. \textbf{125}, no.2, 022301 (2020).
%doi:10.1103/PhysRevLett.125.022301
%[arXiv:1910.14406 [nucl-ex]].
%16 citations counted in INSPIRE as of 05 Dec 2020

%\cite{STAR:2017ckg}
\bibitem{STAR:2017ckg}
L.~Adamczyk \textit{et al.} [STAR],
%``Global $\Lambda$ hyperon polarization in nuclear collisions: evidence for the most vortical fluid,''
Nature \textbf{548}, 62-65 (2017).
%doi:10.1038/nature23004
%[arXiv:1701.06657 [nucl-ex]].
%402 citations counted in INSPIRE as of 21 Nov 2021

%\cite{Muller:2018ibh}
\bibitem{Muller:2018ibh}
B.~M\"uller and A.~Sch\"afer,
%``Chiral magnetic effect and an experimental bound on the late time magnetic field strength,''
Phys. Rev. D \textbf{98}, no.7, 071902 (2018).
%doi:10.1103/PhysRevD.98.071902
%[arXiv:1806.10907 [hep-ph]].
%24 citations counted in INSPIRE as of 21 Nov 2021

%\cite{Guo:2019mgh}
\bibitem{Guo:2019mgh}
X.~Guo, J.~Liao and E.~Wang,
%``Spin Hydrodynamic Generation in the Charged Subatomic Swirl,''
Sci. Rep. \textbf{10}, no.1, 2196 (2020).
%doi:10.1038/s41598-020-59129-6
%[arXiv:1904.04704 [hep-ph]].
%25 citations counted in INSPIRE as of 21 Nov 2021

%\cite{Guo:2019joy}
\bibitem{Guo:2019joy}
Y.~Guo, S.~Shi, S.~Feng and J.~Liao,
%``Magnetic Field Induced Polarization Difference between Hyperons and Anti-hyperons,''
Phys. Lett. B \textbf{798}, 134929 (2019).
%doi:10.1016/j.physletb.2019.134929
%[arXiv:1905.12613 [nucl-th]].
%44 citations counted in INSPIRE as of 21 Nov 2021

%\cite{ALICE:2015mzu}
\bibitem{ALICE:2015mzu}
J.~Adam \textit{et al.} [ALICE],
%``Measurement of an excess in the yield of $J/\psi$ at very low $p_{\rm T}$ in Pb-Pb collisions at $\sqrt{s_{\rm NN}}$ = 2.76 TeV,''
Phys. Rev. Lett. \textbf{116}, no.22, 222301 (2016).
%doi:10.1103/PhysRevLett.116.222301
%[arXiv:1509.08802 [nucl-ex]].
%116 citations counted in INSPIRE as of 21 Nov 2021

%\cite{Zha:2018ytv}
\bibitem{Zha:2018ytv}
W.~Zha, L.~Ruan, Z.~Tang, Z.~Xu and S.~Yang,
%``Coherent photo-produced J$/\psi$ and dielectron yields in isobaric collisions,''
Phys. Lett. B \textbf{789}, 238-242 (2019).
%doi:10.1016/j.physletb.2018.12.041
%[arXiv:1810.02064 [hep-ph]].
%8 citations counted in INSPIRE as of 21 Nov 2021

%\cite{Klein:2018fmp}
\bibitem{Klein:2018fmp}
S.~Klein, A.~H.~Mueller, B.~W.~Xiao and F.~Yuan,
%``Acoplanarity of a Lepton Pair to Probe the Electromagnetic Property of Quark Matter,''
Phys. Rev. Lett. \textbf{122}, no.13, 132301 (2019).
%doi:10.1103/PhysRevLett.122.132301
%[arXiv:1811.05519 [hep-ph]].
%34 citations counted in INSPIRE as of 21 Nov 2021

%\cite{STAR:2019wlg}
\bibitem{STAR:2019wlg}
J.~Adam \textit{et al.} [STAR],
%``Measurement of $e^+e^-$ Momentum and Angular Distributions from Linearly Polarized Photon Collisions,''
Phys. Rev. Lett. \textbf{127}, no.5, 052302 (2021).
%doi:10.1103/PhysRevLett.127.052302
%[arXiv:1910.12400 [nucl-ex]].
%28 citations counted in INSPIRE as of 21 Nov 2021

%\cite{Brandenburg:2021lnj}
\bibitem{Brandenburg:2021lnj}
J.~D.~Brandenburg, W.~Zha and Z.~Xu,
%``Mapping the electromagnetic fields of heavy-ion collisions with the Breit-Wheeler process,''
Eur. Phys. J. A \textbf{57}, no.10, 299 (2021).
%doi:10.1140/epja/s10050-021-00595-5
%[arXiv:2103.16623 [hep-ph]].
%5 citations counted in INSPIRE as of 21 Nov 2021

%\cite{Gursoy:2014aka}
\bibitem{Gursoy:2014aka}
U.~Gursoy, D.~Kharzeev and K.~Rajagopal,
%``Magnetohydrodynamics, charged currents and directed flow in heavy ion collisions,''
Phys. Rev. C \textbf{89}, 054905 (2014).
%doi:10.1103/PhysRevC.89.054905
%[arXiv:1401.3805 [hep-ph]].
%162 citations counted in INSPIRE as of 07 Dec 2020

%\cite{Tuchin:2015oka}
\bibitem{Tuchin:2015oka}
K.~Tuchin,
%``Initial value problem for magnetic fields in heavy ion collisions,''
Phys. Rev. C \textbf{93}, 014905 (2016).
%doi:10.1103/PhysRevC.93.014905
%[arXiv:1508.06925 [hep-ph]].
%50 citations counted in INSPIRE as of 08 Dec 2020

%\cite{Zakharov:2014dia}
\bibitem{Zakharov:2014dia}
B.~G.~Zakharov,
%``Electromagnetic response of quark\textendash{}gluon plasma in heavy-ion collisions,''
Phys. Lett. B \textbf{737}, 262-266 (2014).
%doi:10.1016/j.physletb.2014.08.068
%[arXiv:1404.5047 [hep-ph]].
%51 citations counted in INSPIRE as of 08 Dec 2020

  
%\cite{Pu:2016ayh}
\bibitem{Pu:2016ayh}
S.~Pu, V.~Roy, L.~Rezzolla and D.~H.~Rischke,
%``Bjorken flow in one-dimensional relativistic magnetohydrodynamics with magnetization,''
Phys. Rev. D \textbf{93}, 074022 (2016).
%doi:10.1103/PhysRevD.93.074022
%[arXiv:1602.04953 [nucl-th]].
%48 citations counted in INSPIRE as of 07 Dec 2020


%\cite{Inghirami:2016iru}
\bibitem{Inghirami:2016iru} 
  G.~Inghirami, L.~Del Zanna, A.~Beraudo, M.~H.~Moghaddam, F.~Becattini and M.~Bleicher,
  %``Numerical magneto-hydrodynamics for relativistic nuclear collisions,''
  Eur.\ Phys.\ J.\ C {\bf 76}, 659 (2016). 

%\cite{Yan:2021zjc}
\bibitem{Yan:2021zjc}
L.~Yan and X.~G.~Huang,
%``Dynamical evolution of magnetic field in the pre-equilibrium quark-gluon plasma,''
arXiv:2104.00831 [nucl-th].
%6 citations counted in INSPIRE as of 19 Oct 2021

%\cite{Stewart:2017zsu}
\bibitem{Stewart:2017zsu}
E.~Stewart and K.~Tuchin,
%``Magnetic field in expanding quark-gluon plasma,''
Phys. Rev. C \textbf{97}, no.4, 044906 (2018).
%doi:10.1103/PhysRevC.97.044906
%[arXiv:1710.08793 [nucl-th]].
%17 citations counted in INSPIRE as of 21 Nov 2021

%\cite{Stewart:2021mjz}
\bibitem{Stewart:2021mjz}
E.~Stewart and K.~Tuchin,
%``Continuous evolution of electromagnetic field in heavy-ion collisions,''
Nucl. Phys. A \textbf{1016}, 122308 (2021).
%doi:10.1016/j.nuclphysa.2021.122308
%[arXiv:2106.09124 [nucl-th]].
%2 citations counted in INSPIRE as of 21 Nov 2021

%\cite{Chen:2021nxs}
\bibitem{Chen:2021nxs}
Y.~Chen, X.~L.~Sheng and G.~L.~Ma,
%``Electromagnetic fields from the extended Kharzeev-McLerran-Warringa model in relativistic heavy-ion collisions,''
Nucl. Phys. A \textbf{1011}, 122199 (2021).
%doi:10.1016/j.nuclphysa.2021.122199
%[arXiv:2101.09845 [nucl-th]].
%1 citations counted in INSPIRE as of 28 Nov 2021

%\cite{Hattori:2016emy}
\bibitem{Hattori:2016emy}
K.~Hattori and X.~G.~Huang,
%``Novel quantum phenomena induced by strong magnetic fields in heavy-ion collisions,''
Nucl. Sci. Tech. \textbf{28}, no.2, 26 (2017).
%doi:10.1007/s41365-016-0178-3
%[arXiv:1609.00747 [nucl-th]].
%98 citations counted in INSPIRE as of 28 Nov 2021

%\cite{Schenke:2010nt}
\bibitem{Schenke:2010nt}
B.~Schenke, S.~Jeon and C.~Gale,
%``(3+1)D hydrodynamic simulation of relativistic heavy-ion collisions,''
Phys. Rev. C \textbf{82}, 014903 (2010).
%doi:10.1103/PhysRevC.82.014903
%[arXiv:1004.1408 [hep-ph]].
%335 citations counted in INSPIRE as of 28 Nov 2021

%\cite{Kestin:2008bh}
\bibitem{Kestin:2008bh}
G.~Kestin and U.~W.~Heinz,
%``Hydrodynamic radial and elliptic flow in heavy-ion collisions from AGS to LHC energies,''
Eur. Phys. J. C \textbf{61}, 545-552 (2009).
%doi:10.1140/epjc/s10052-008-0832-y
%[arXiv:0806.4539 [nucl-th]].
%57 citations counted in INSPIRE as of 28 Nov 2021
  
%\cite{Arnold:2003zc}
\bibitem{Arnold:2003zc}
P.~B.~Arnold, G.~D.~Moore and L.~G.~Yaffe,
%``Transport coefficients in high temperature gauge theories. 2. Beyond leading log,''
JHEP \textbf{05}, 051 (2003).
%doi:10.1088/1126-6708/2003/05/051
%[arXiv:hep-ph/0302165 [hep-ph]].
%624 citations counted in INSPIRE as of 08 Jun 2021  
  
%\cite{Gupta:2003zh}
\bibitem{Gupta:2003zh}
S.~Gupta,
%``The Electrical conductivity and soft photon emissivity of the QCD plasma,''
Phys. Lett. B \textbf{597}, 57-62 (2004).
%doi:10.1016/j.physletb.2004.05.079
%[arXiv:hep-lat/0301006 [hep-lat]].
%174 citations counted in INSPIRE as of 08 Jun 2021  

%\cite{Aarts:2007wj}
\bibitem{Aarts:2007wj} 
  G.~Aarts, C.~Allton, J.~Foley, S.~Hands and S.~Kim,
  %``Spectral functions at small energies and the electrical conductivity in hot, quenched lattice QCD,''
  Phys.\ Rev.\ Lett.\  {\bf 99}, 022002 (2007).  

%\cite{Buividovich:2010tn}
\bibitem{Buividovich:2010tn} 
  P.~V.~Buividovich, M.~N.~Chernodub, D.~E.~Kharzeev, T.~Kalaydzhyan, E.~V.~Luschevskaya and M.~I.~Polikarpov,
  %``Magnetic-Field-Induced insulator-conductor transition in SU(2) quenched lattice gauge theory,''
  Phys.\ Rev.\ Lett.\  {\bf 105}, 132001 (2010).

%\cite{Ding:2010ga}
\bibitem{Ding:2010ga} 
  H.-T.~Ding, A.~Francis, O.~Kaczmarek, F.~Karsch, E.~Laermann and W.~Soeldner,
  %``Thermal dilepton rate and electrical conductivity: An analysis of vector current correlation functions in quenched lattice QCD,''
  Phys.\ Rev.\ D {\bf 83}, 034504 (2011).
  
%\cite{Burnier:2012ts}
\bibitem{Burnier:2012ts} 
  Y.~Burnier and M.~Laine,
  %``Towards flavour diffusion coefficient and electrical conductivity without ultraviolet contamination,''
  Eur.\ Phys.\ J.\ C {\bf 72}, 1902 (2012).  
  
%\cite{Brandt:2012jc}
\bibitem{Brandt:2012jc} 
  B.~B.~Brandt, A.~Francis, H.~B.~Meyer and H.~Wittig,
  %``Thermal Correlators in the \rho\ channel of two-flavor QCD,''
  JHEP {\bf 03}, 100 (2013).
  
%\cite{Amato:2013naa}
\bibitem{Amato:2013naa} 
  A.~Amato, G.~Aarts, C.~Allton, P.~Giudice, S.~Hands and J.~I.~Skullerud,
  %``Electrical conductivity of the quark-gluon plasma across the deconfinement transition,''
  Phys.\ Rev.\ Lett.\  {\bf 111}, 172001 (2013).    
  
%\cite{Finazzo:2013efa}
\bibitem{Finazzo:2013efa} 
  S.~I.~Finazzo and J.~Noronha,
  %``Holographic calculation of the electric conductivity of the strongly coupled quark-gluon plasma near the deconfinement transition,''
  Phys.\ Rev.\ D {\bf 89}, 106008 (2014).

%\cite{Sahoo:2018dxn}
\bibitem{Sahoo:2018dxn} 
  P.~Sahoo, S.~K.~Tiwari and R.~Sahoo,
  %``Electrical conductivity of hot and dense QCD matter created in heavy-ion collisions: A color string percolation approach,''
  Phys.\ Rev.\ D {\bf 98}, 054005 (2018).    

%\cite{Cassing:2013iz}
\bibitem{Cassing:2013iz} 
  W.~Cassing, O.~Linnyk, T.~Steinert and V.~Ozvenchuk,
  %``Electrical Conductivity of Hot QCD Matter,''
  Phys.\ Rev.\ Lett.\  {\bf 110}, 182301 (2013).
  
%\cite{Greif:2014oia}
\bibitem{Greif:2014oia} 
  M.~Greif, I.~Bouras, C.~Greiner and Z.~Xu,
  %``Electric conductivity of the quark-gluon plasma investigated using a perturbative QCD based parton cascade,''
  Phys.\ Rev.\ D {\bf 90}, 094014 (2014).  
  
%\cite{Puglisi:2014sha}
\bibitem{Puglisi:2014sha} 
  A.~Puglisi, S.~Plumari and V.~Greco,
  %``Electric Conductivity from the solution of the Relativistic Boltzmann Equation,''
  Phys.\ Rev.\ D {\bf 90}, 114009 (2014). 

%\cite{Steinert:2013fza}
\bibitem{Steinert:2013fza}
T.~Steinert and W.~Cassing,
%``Electric and magnetic response of hot QCD matter,''
Phys. Rev. C \textbf{89}, no.3, 035203 (2014).
%doi:10.1103/PhysRevC.89.035203
%[arXiv:1312.3189 [hep-ph]].
%60 citations counted in INSPIRE as of 03 Jun 2021

%\cite{Xu:2004mz}
\bibitem{Xu:2004mz}
Z.~Xu and C.~Greiner,
%``Thermalization of gluons in ultrarelativistic heavy ion collisions by including three-body interactions in a parton cascade,''
Phys. Rev. C \textbf{71}, 064901 (2005).
%doi:10.1103/PhysRevC.71.064901
%[arXiv:hep-ph/0406278 [hep-ph]].
%452 citations counted in INSPIRE as of 05 Dec 2020  

%\cite{Greif:2017irh}
\bibitem{Greif:2017irh}
M.~Greif, C.~Greiner and Z.~Xu,
%``Magnetic field influence on the early time dynamics of heavy-ion collisions,''
Phys. Rev. C \textbf{96}, no.1, 014903 (2017).
%doi:10.1103/PhysRevC.96.014903
%[arXiv:1704.06505 [hep-ph]].
%17 citations counted in INSPIRE as of 21 Feb 2022

%\cite{Xu:2007aa}
\bibitem{Xu:2007aa}
Z.~Xu and C.~Greiner,
%``Transport rates and momentum isotropization of gluon matter in ultrarelativistic heavy-ion collisions,''
Phys. Rev. C \textbf{76}, 024911 (2007).
%doi:10.1103/PhysRevC.76.024911
%[arXiv:hep-ph/0703233 [hep-ph]].
%147 citations counted in INSPIRE as of 28 Apr 2021

%\cite{Hirano:2005wx}
\bibitem{Hirano:2005wx}
T.~Hirano and M.~Gyulassy,
%``Perfect fluidity of the quark gluon plasma core as seen through its dissipative hadronic corona,''
Nucl. Phys. A \textbf{769}, 71-94 (2006).
%doi:10.1016/j.nuclphysa.2006.02.005
%[arXiv:nucl-th/0506049 [nucl-th]].
%333 citations counted in INSPIRE as of 21 Oct 2021

%\cite{El:2012cr}
\bibitem{El:2012cr}
A.~El, F.~Lauciello, C.~Wesp, Z.~Xu and C.~Greiner,
%``Shear viscosity of an ultrarelativistic Boltzmann gas with isotropic inelastic scattering processes,''
Nucl. Phys. A \textbf{925}, 150-160 (2014).
%doi:10.1016/j.nuclphysa.2014.02.009
%[arXiv:1207.5331 [hep-th]].
%9 citations counted in INSPIRE as of 21 Oct 2021

%\cite{Blaizot:2011xf}
\bibitem{Blaizot:2011xf}
J.~P.~Blaizot, F.~Gelis, J.~F.~Liao, L.~McLerran and R.~Venugopalan,
%``Bose--Einstein Condensation and Thermalization of the Quark Gluon Plasma,''
Nucl. Phys. A \textbf{873}, 68-80 (2012).
%doi:10.1016/j.nuclphysa.2011.10.005
%[arXiv:1107.5296 [hep-ph]].
%173 citations counted in INSPIRE as of 18 May 2021

%\cite{Romatschke:2017vte}
\bibitem{Romatschke:2017vte}
P.~Romatschke,
%``Relativistic Fluid Dynamics Far From Local Equilibrium,''
Phys. Rev. Lett. \textbf{120}, 012301 (2018).
%doi:10.1103/PhysRevLett.120.012301
%[arXiv:1704.08699 [hep-th]].
%100 citations counted in INSPIRE as of 09 Dec 2020



\end{thebibliography}
\end{document}